\documentclass[conference,compsoc]{IEEEtran}
\ifCLASSOPTIONcompsoc
  \usepackage[nocompress]{cite}
\else
  \usepackage{cite}
\fi
\ifCLASSINFOpdf
\else
\fi
\usepackage{url}
\usepackage{tikz}
\usepackage{amsmath}

\usepackage{xcolor}
\usepackage{subcaption}
\usepackage{listings}
\usepackage{framed}
\usepackage{fancyhdr}
\usepackage{tabularx}
\usepackage{adjustbox}
\usepackage[normalem]{ulem}
\usepackage{graphicx}
\usepackage{booktabs}
\usepackage{enumitem}
\usepackage[plainpages=false]{hyperref}

\lstdefinestyle{verilog-style}
{
    language=Verilog,
    basicstyle=\small\ttfamily,
    keywordstyle=\color{black},
    identifierstyle=\color{black},
    commentstyle=\color{black},
    numbers=left,
    basicstyle=\small,
    numberstyle=\tiny\color{black},
    numbersep=10pt,
    tabsize=8,
    literate=*{:}{:}1
}

\hypersetup{ colorlinks = true, %Colours links instead of ugly boxes 
urlcolor = blue, %Colour for external hyperlinks 
linkcolor = blue, %Colour of internal links 
citecolor = blue %Colour of citations 
}

\begin{document}
\title{Augmented Symbolic Execution for Information Flow in Hardware
  Designs}

\author{\IEEEauthorblockN{Anon}}
%% \IEEEauthorblockA{School of Electrical and\\Computer Engineering\\
%% Georgia Institute of Technology\\
%% Atlanta, Georgia 30332--0250\\
%% Email: http://www.michaelshell.org/contact.html}
%% \and
%% \IEEEauthorblockN{Homer Simpson}
%% \IEEEauthorblockA{Twentieth Century Fox\\
%% Springfield, USA\\
%% Email: homer@thesimpsons.com}
%% \and
%% \IEEEauthorblockN{James Kirk\\ and Montgomery Scott}
%% \IEEEauthorblockA{Starfleet Academy\\
%% San Francisco, California 96678-2391\\
%% Telephone: (800) 555--1212\\
%% Fax: (888) 555--1212}}


% compsoc conferences compared to traditional conferences), use this

\author{\IEEEauthorblockN{Kaki Ryan} 
University of North Carolina \\
Chapel Hill, NC, USA
\and
\IEEEauthorblockN{Matthew Gregoire}
University of North Carolina \\
Chapel Hill, NC, USA
\and
\IEEEauthorblockN{Cynthia Sturton} 
University of North Carolina \\
Chapel Hill, NC, USA
}


\maketitle

\begin{abstract}
We present \emph{SEIF}, a methodology that combines static analysis with
symbolic execution to verify and explicate information flow paths in a hardware
design. SEIF begins with a statically built model of the information flow through a
design and uses guided symbolic execution to recognize and eliminate
non-flows with high precision or to find corresponding paths through the design
state for true flows. We evaluate SEIF on two open-source CPUs, an AES core, and the
AKER access control module. SEIF can exhaustively explore 10-12 clock cycles deep in
4-6 seconds on average, and can automatically account for 86-90\% of 
the paths in the statically built model. %% , verifying 58-77\% of them as repayable from the design's
%% reset state.
Additionally, SEIF can be used to find multiple violating paths for security
properties, providing a new angle for security verification.

%10 pages, excluding refs and appdx
\end{abstract}


% \fi
\IEEEpeerreviewmaketitle

\section{Introduction}
%% This paper presents \textit{SEIF} (pronounced ``safe''), a toolflow that combines static analysis with symbolic execution
%%  for verifying information flow paths in hardware designs. SEIF begins with a static model of the information flows
%%  through a design and is able to produce realizable and semantically accurate information flow paths. This is made
%%  possible by an augmented symbolic execution engine with heuristics that steer the search towards paths
%%  necessary to complete the flow. The search space would otherwise be intractable.
%%  SEIF is able to produce counterexamples when given a security property, in addition 
%%  to providing all of the replayable information flow paths that exhibit the vulnerable behavior.
 
%% Analyzing information flow is important for security
Analyzing how information flows through a hardware design is critical
to verifying the security of the design~\cite{tiwari2009complete,tiwari2009execution,jin2012proof,li2011caisson,li2014sapper,zhang2015secverilog, bidmeshki2015vericoq, hu2016detecting,kong2017using,ardeshiricham2017register,ardeshiricham2017clepsydra,deng2017secchisel,bidmeshki2017information,boraten2018securing,pilato2018tainthls,zagieboylo2019using,pieper2020dynamic}.
Unwanted flows of information to
or from a signal in the design can violate desired security policies in the form of
 access authorization
violations~\cite{restuccia2021aker,restuccia2022framework}, memory leakage
vulnerabilities~\cite{cherupalli2017software,fadiheh2023exhaustive}, %% \cks{smthg for Isadora
  %% paper~\cite{Deutschbein2021Isadora,Deutschbein2022JCEN}, Chapter in
  %% txtbk~\cite{sturton2022handbook}, Kastner's survey paper on IF Verification~\cite{??}},
and possible privilege escalation vulnerabilities~\cite{wu2022exert}.
Recent work has shown that symbolic execution
is a powerful tool for analyzing how information flows through a hardware
design~\cite{athalye2022knox,fowze2022eisec,cherupalli2017software}. Symbolic analysis is
precise and enables
tracking both direct and indirect flows of information through each path of
execution without instrumenting the design with added tracking logic.

%% Current practice for detecting and
%% understanding these flows uses simulation-based testing~\cite{smth}, static
%% analysis~\cite{smth}, or most recently, symbolic execution~\cite{smth}.

%% Still not a solved problem
Unfortunately, symbolic execution infamously suffers from the path explosion
problem. The number of paths through a design grows exponentially with the
number of branch points in the design. Hardware designs have the added
complexity of reasoning about paths over multiple clock cycles in order to
realize complete flows of information from an input port (source) signal to an output
port (sink) signal. Current solutions to the path explosion problem have been to consider
small, but security critical designs~\cite{athalye2022knox,fowze2022eisec}, or to constrain the hardware design
space by analyzing how information flows for a particular software program~\cite{cherupalli2017software,athalye2019notary}.

%% Static analysis plus symbolic execution is the answer!
We take a different approach. We start with static analysis, using an existing
tool~\cite{meza2023hyperflowgraph} to
build a graph that over-approximates how information flows
through a design. Such a graph is useful for designers, allowing them to explore
their design and find possible illegal or insecure flows.  %% We show that
%% by using the information in this graph to guide the symbolic execution engine, we
%% can fully explore how information flows from an input port signal of interest through
%% a design for a sufficient number of clock cycles for the flow to reach all possible
%% output ports. The approach can scale to full CPU designs. (In our experiments this is typically \cks{XX-XX} clock cycles for
%% an open-source CPU~\ref{evaluation}.)
For our purposes, the graph provides useful information that can be used to guide
symbolic execution: a sequence of landmark points in
the hardware design that execution must reach in order to
realize a given path of information flow. Using these landmarks as a guide, we use symbolic
execution to improve the graph's efficacy: finding a realizable path through the
design state, along with the inputs needed to take that path, corresponding to a
path in the graph; recognizing and eliminating from the graph paths which are
unrealizable in execution; and recognizing and eliminating from the graph paths
which are realizable, but do not represent a true flow of information.

%% some paths as unrealizable, and for the paths remaining, develop
%% heuristics that steer
%% symbolic execution toward paths more likely to result in the desired information flow.

%% But the graph does not tell us which paths
%% of execution to take in order to get to each subsequent in landmark when we are looking
%% at a path made up of a sequence of many information flows. It also doesn't tell us 
%% how long (how many clock cycles of execution) it will take to get there, or whether it is even
%% possible to get to the next landmark at all from current one. To answer these
%% questions we use develop and implement search heuristics based on SMT solving into a symbolic execution
%% engine. \cks{This is a
%%   reachability problem. Piecewise reachability? Stepwise reachability?}
%%  %stalling, unsat core?, backtracking.

This paper presents \textit{SEIF} (pronounced ``safe''), a toolflow that
combines symbolic execution with static analysis in the form of the information
flow graph. SEIF takes as input the statically built
information flow (IF) graph and the source signals of interest in the design.
Three
outcomes are possible: 1. SEIF finds that the path is unrealizable or does not
represent a true flow of information, and requires no
further scrutiny from the security engineer, 2. SEIF returns a sequence of input
values that will drive the design along the IF-graph path to realize the flow of
information, or 3. the complexity of the search space
leaves the IF-graph path unaccounted for.

To find that a path is unrealizable or does not represent a true flow of
information, SEIF uses two mechanisms: a check for mutually contradictory constraints and symbolic
analysis. If the first mechanism reports that a path is unrealizable then it is, regardless of the
number of clock cycles the design is allowed to run. If the second mechanism
reports that a path is unrealizable, then it is within the clock cycle bound used
by SEIF. In our experiments, this was the case for 5-7\% of the paths. 

To find and return a sequence of input values that will drive execution along an
IF-graph path, SEIF uses symbolic execution guided by the IF graph and
heuristics we develop. The returned
sequence of input values will drive execution along the IF-graph path, either
starting from the design's reset state or from an intermediate state. In our
evaluation (Section~\ref{sec:eval}) we differentiate these two cases.

In this paper, we develop SEIF, an algorithm and tool to search for and eliminate
false paths of information flow from a static analysis of a hardware
design and then to further explicate the paths that remain. We show that by
using the static analysis as a guide, we can guide symbolic execution toward more probable paths and eliminate impossible paths
early. Our contributions are:
\begin{itemize}
\item Define \emph{SEIF}, an augmented symbolic execution methodology for information flow
  analysis. %% Develop the use of symbolic execution with unsat core
  %% evaluation to efficiently remove false positive paths from a static information
  %% flow analysis and enable more dynamic security analysis
\item Implement the methodology and search heuristics on top of the symbolic execution engine discussed in \cite{ryan2023countering}.
\item Evaluate the augmented symbolic execution strategy on four open-source designs.
\end{itemize}

\section{Threat Model}
Information-flow analysis is a part of the security validation
activities~\cite{dorsey2020intel} that take place during the design phase of
the hardware lifecycle~\cite{he2015model}. The goal is to find
weaknesses, vulnerabilities, and flaws in register transfer level (RTL)
designs that may be exploitable post-deployment.

%% Our threat model is an attacker, or colluding attackers, acting in the
%% post-deployment phase, after the hardware has been fabricated and shipped. The
%% attackers aim to exploit a hardware vulnerability to gain unauthorized access to
%% data or execution. For example, a hardware flaw may allow the attacker to learn
%% confidential data stored in memory or hardcoded into a device, or it may allow
%% the attacker to modify the control flow of software running on a processor by
%% leveraging aspects of the processor's behavior. The attacker may access the
%% hardware locally. Remote attacks caused by running attacker-controlled software
%% (e.g., JavaScript on a website visited by the machine) are feasible. Physical
%% attacks, including hardware tampering, fault injection, and power or
%% electromagnetic side channels, are out of scope.
Flaws that result from logic and physical synthesis
tools, manufacturing, or the supply chain cannot be discovered by SEIF. 
We target flaws that occur by benign human error in the specification,
design, or implementation phases. Our analysis may find maliciously
inserted flaws, they will have a lower chance of being uncovered than benign
flaws as the attacker will likely take steps to hide their work,
so that the security engineer does not recognize the
malicious flow of information as dangerous. Flaws maliciously inserted after the
security validation is complete, e.g., analog Trojans~\cite{YangSP2016}, cannot
be discovered by SEIF.

%% Finally, the security engineer who is conducting the information-flow analysis
%% is trusted to act in good faith. The goal of our tool is to help the engineer
%% pare down the set of possible flows of information requiring study, and provide
%% concrete test cases for true flows of information to help the engineer better
%% understand the design. This tool supports the imperfect engineer by helping them
%% to understand the design better and highlight the security implications of
%% various information flows.

\section{Problem Statement}

We approach the problem of information-flow analysis by
transforming it into a graph reachability problem over a labeled, directed graph
representing signal connectivity, extracted from the Verilog RTL design. We use symbolic execution of the RTL
to determine which paths through the labeled directed graph represent true flows of information through
the design in execution. 

%We can also think of this as using symbolic execution to model how
%information flows through the design, and using the static signal connectivity graph
%to guide the symbolic execution engine through the
%large search space.

%The signal connectivity graph helps us tackle
%the path explosion problem inherent in symbolic execution. 

%Our problem statement is as follows.
%Given a hardware design at the register transfer level (RTL), and an input
%signal of interest, $x$, determine the information
%flow of that signal. Information flows from signal $x$ to signal $y$ when the value
%of $y$ at any clock cycle depends on or is influenced by the value of $x$. 

%This
%definition allows for checking
%non-interference~\cite{goguen1982security}: under this definition, we can say
%that $x$ is non-interfering with
% $y$ if there exist no execution traces
%in which $x$ flows to $y$.

Given a hardware design and a particular input signal of
interest, the goal is to return:
\begin{enumerate}[noitemsep]
\item  the set of realizable information flows through the design originating at
  that signal; and
\item for each found information flow, return a sequence of input values to the design that will
  drive the information flow.
%% \item \cks{OR: (Kaki, which do you think is more accurate?)} 
%% \item the set of variables in the design that are influenced by the input
%%   signal $v$; and 
%% \item for each found signal, return a sequence of input values to the design
%%   that will drive the information flow from $v$ to that signal.
\end{enumerate}

\section{Preliminaries}
It is useful to keep in
mind three models: the state diagram of the design
showing machine states and transitions between them; the labeled, directed, signal-connectivity graph, which we
call the \emph{Information Flow (IF)} graph; and the symbolic execution
(SE) tree, showing execution paths through the RTL along with the associated
(symbolic) states and path conditions. We describe these three in the following
sections, but first we introduce a fragment of Verilog RTL as a toy example to
help illustrate the three models.

\subsection{Toy Example}
The code snippet of Figure~\ref{fig:toy1} shows a flow from an input, \texttt{secret}, to an output, \texttt{led}. The flow is
guarded by an internal, state-holding variable and the secret will only flow to the
LED output in the clock cycle after $\mathtt{state} = 3$. Note that with non-blocking assignments (``$<=$'')
 all right-hand side expressions are calculated at the same time and assignments
 take effect at the next clock cycle. Blocking assignments (``$=$'') take effect
 immediately.

  \begin{figure}[h!]
  \centering
  \begin{framed}
    \begin{lstlisting}[style={verilog-style},belowskip=-.8\baselineskip,aboveskip=-.5\baselineskip]
always @(posedge clk) begin
   if (enable) begin
      prev <= state;
      state <= state + 1;
   end

   if (state == 3) begin
      guard <= secret;
   end else begin
      guard <= 0;
   end
    
end
assign led = (prev == 3) ? guard : 0;
    \end{lstlisting}
  \end{framed}
  \caption{Toy example. \texttt{clk}, \texttt{enable}, and \texttt{secret} are input \textbf{wire}s.
\texttt{state}, \texttt{prev}, and \texttt{guard} are state-holding \textbf{reg}s. Not
shown is the initialization, which sets \texttt{state}, \texttt{prev}, and \texttt{guard} to 0.
\texttt{led} is an output wire. \texttt{secret} flows through \texttt{guard} to
    \texttt{led} after four clock cycles.}
  \label{fig:toy1}
  \end{figure}

  \subsection{State Diagram}

We model a
hardware design as a tuple, $D = (S, s_0, I, \delta, \omega)$,
where
\begin{itemize}[noitemsep]
\item $S$ is the set of states of the design;
\item $S_0 \subset S$ is the set of initial states;
\item $I$ is the finite set of input strings;
\item $\delta: S \times I \rightarrow S$ is the transition function;
\item $\omega: S \rightarrow O$ is the output function.
\end{itemize}

A state $s \in S$ is a vector of valuations to state-holding internal variables of the design,
$s = \langle v_0, v_1, \ldots, v_{|s|}\rangle$. We use $v_i$
to indicate the variable and $\langle v_i = x \rangle_{s_j}$ to indicate that the value of
variable $v_i$ is $x$ in state $s_j$. As shorthand, we sometimes use $v_i$ to
refer to both the variable and its value, when it is clear in the text what we
mean. The design powers up in an
initial state, $s_0$. Many state-holding variables are reset to 0 in the initial
state. An input string $i \in I$ is a concatenation of values to input variables
of the design. Inputs are provided on every clock cycle. Similarly to
state-holding variables, we refer to the value of input variable $v_j$ at any given clock
cycle as $\langle v_j = x \rangle$ or simply $v_j$. The
\texttt{clk} signal is a special input that synchronizes reading input values and state
transitions, which happen on clock cycle edges. The output function is the
identity function over a subset of the design's variables.

\begin{figure}[h]
  \centering
    \setlength{\belowcaptionskip}{-10pt}
  \includegraphics[width=\columnwidth,trim = 15 225 30 125, clip]{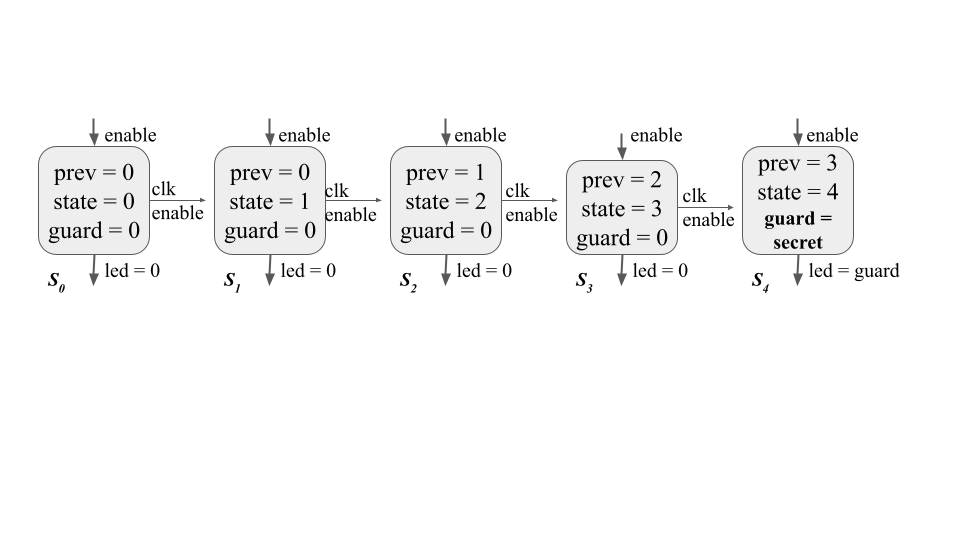}
  \caption{State transitions of the toy example (Figure~\ref{fig:toy1}) in which information flows from
    \texttt{secret} to \texttt{led}.}
  \label{fig:toy1states}
\end{figure}

For example, Figure~\ref{fig:toy1states} shows a sequence of state transitions for the toy example, starting
with the initial state, in which
information flows from
 \texttt{secret} to \texttt{guard} to output variable \texttt{led}.
In this example, the initial state $s_0 = \langle
\texttt{prev} = 0, \texttt{state} = 0, \texttt{guard} = 0\rangle$ produces output
$\omega(s_0) = \langle\texttt{led} = 0\rangle$, and transitions to state $s_1 = \langle\texttt{prev} = 0,
\texttt{state} = 1, \texttt{guard} = 0\rangle$ when \texttt{enable} is high on a positive clock edge.

 \subsection{Information Flow (IF) Graph}

\begin{figure}[h]
  \centering
    \setlength{\belowcaptionskip}{-5pt}
  \includegraphics[scale=0.20]{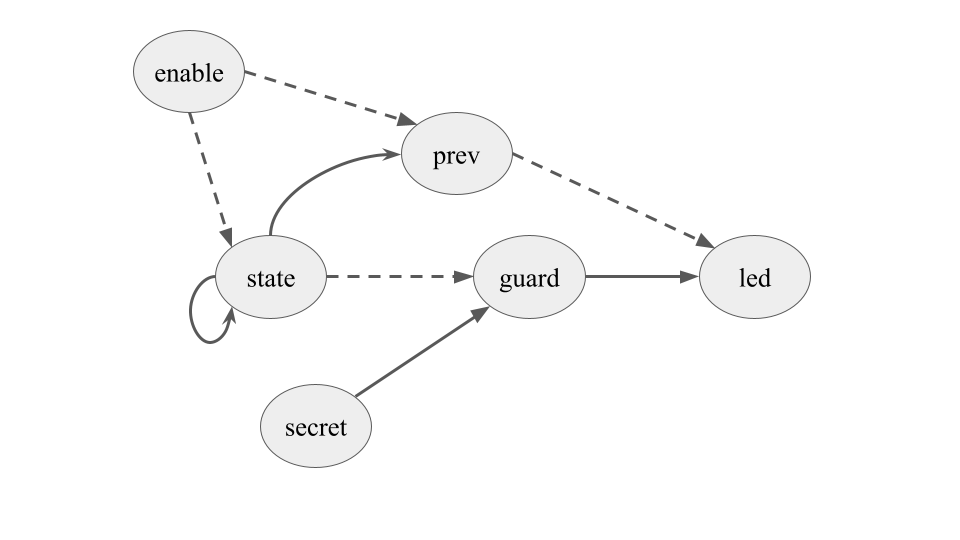}
  \caption{An IF graph for the code in Figure~\ref{fig:toy1}. Dashed lines represent implicit flows
    of information and solid lines represent explicit flows. Labels are
  omitted for space.} 
  \label{fig:toy1ifg}
\end{figure}

The Information Flow (IF) graph is a labeled, directed graph that captures signal
connectivity and provides additional information, taken from the Verilog source,
about the conditions under which two signals are connected~\cite{meza2023hyperflowgraph}. Nodes represent the variables (\texttt{wire}s
and \texttt{reg}s) of the design, and edges indicate a possible flow of
information from one variable to another. An edge $(v_1,v_2)$
exists when either there is an assignment from $v_1$ to $v_2$ (e.g., $v_2 <=
v_1$) or $v_1$ appears in a condition (e.g., $\mathtt{if} (v_1)$), and $v_2$ appears
on the left-hand side of an assignment in either branch. The edge is labeled
with the line number of the relevant Verilog statement and lists the surrounding
conditions in the code that must be true for the information flow to take
place. For example, in Figure~\ref{fig:toy1ifg}, which shows the IF graph for the code in
Figure~\ref{fig:toy1}, the edge $(\mathtt{secret}, \mathtt{guard})$ would be
labeled with the condition that $\mathtt{state} == 3$. Note that this graph inherently has no notion of timing or
clock cycles.

Each edge in the IF graph represents a viable 1-hop flow of
information in the design. However, multi-hop paths through the IF graph may not
correspond to viable information flows. In other words, if
we view the IF graph as an information-flow relation, taking the
transitive closure of the relation yields an over-approximation of information
flow through the design. To demonstrate, consider the code in
Figure~\ref{fig:toy1}, but with the last line replaced with the following:
\begin{verbatim}
assign led = (prev == 2) ? guard : 0;
\end{verbatim} 

The IF graph would have the same nodes and edges, but the path from
\texttt{secret} to \texttt{guard} to \texttt{led} does not correspond to any
flow of information through the design.

There are two reasons why a path through the IF graph may not correspond to a
true information flow. The first is that, as in the example above,
the sequence of conditions needed for
each edge cannot be satisfied. The second is that a path through the IF graph
from $x$ to $y$ may not correspond to a true flow of
information in the sense that the value of $y$ depends on the value of $x$. A
common example of this is the assignment $y = x \oplus x$.

%We use symbolic execution to find which
%paths through the IF graph represent true flows of information through the design.

\subsection{Symbolic Execution}
In symbolic execution, concrete input values are replaced with abstract
symbols. The design is executed using the symbols in place of
literals. When a branch point (e.g., $\mathtt{if}$\texttt{(enable)}) is reached, both paths are separately
explored. For each path, the branching condition that must be true for that path
(e.g., \texttt{enable} $== 1$).
is maintained in the
\emph{path condition}. At the end of a single path of symbolic execution, satisfying
assignments to the constraints in the path condition can be used as concrete
input values to drive concrete execution down that same path.

Symbolic execution is modeled as a directed tree. Each node $n$ in the tree
is associated with a line of code in the design and is associated with a symbolic state, $\sigma$, and path condition, $\pi$. A node's children are the possible next
lines of code to symbolically execute. A path from the root node to any leaf
node represents a realizable path through the design.

\begin{figure}[h]
  \centering
  \includegraphics[width=\columnwidth,trim = 0 5 0 5, clip]{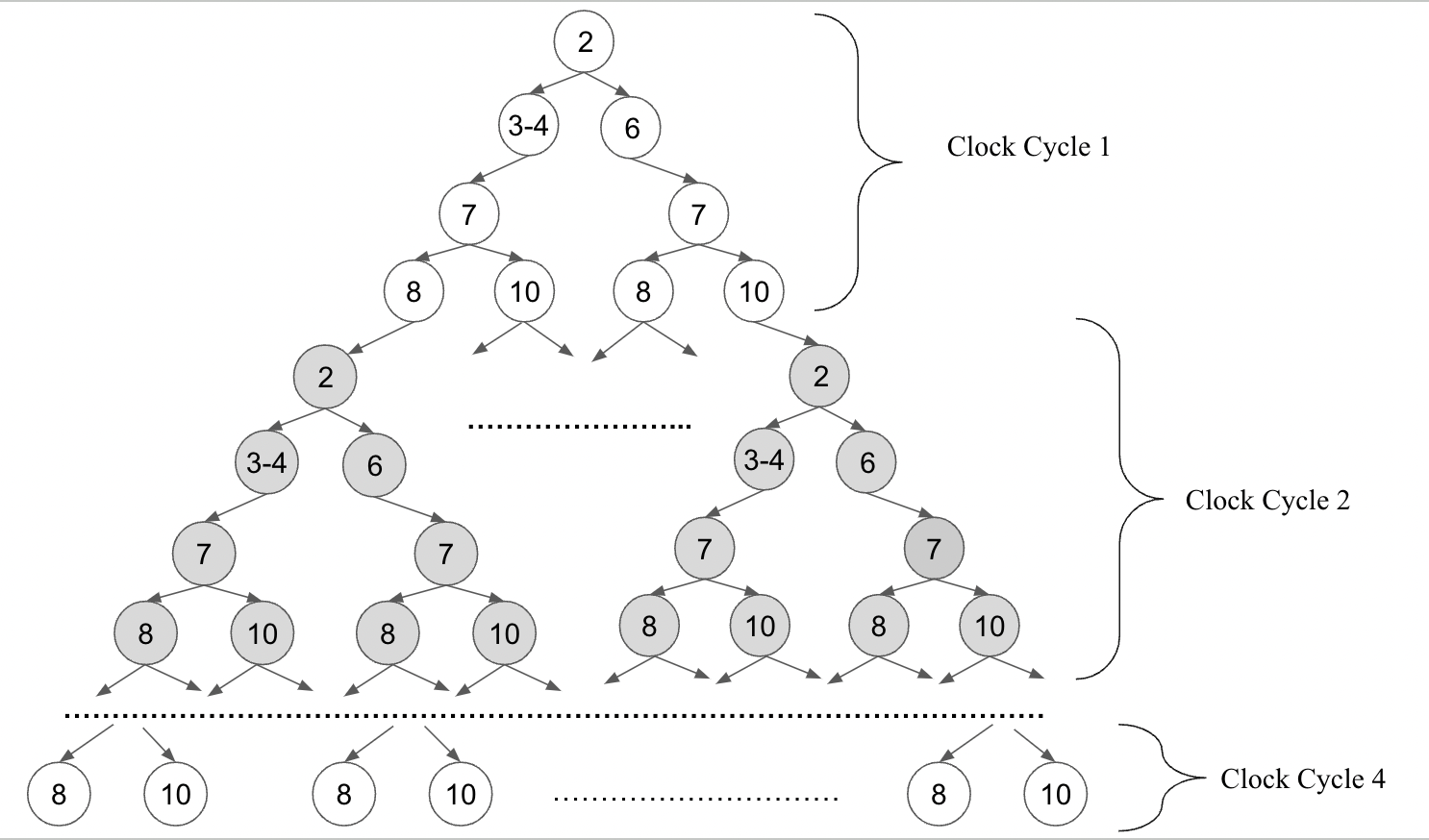}
  \caption{Symbolic execution tree of the design in Figure~\ref{fig:toy1} after
    four clock cycles.}
  \label{fig:toy1tree}
\end{figure}

The number of paths to explore grows quickly. For example, the symbolic execution of the design in Figure~\ref{fig:toy1} for
the four clock cycles necessary to find the information-flow path from
\texttt{secret} to \texttt{led} would yield the tree of nodes shown in Figure~\ref{fig:toy1tree}.
%%
%%
%% single clock cycle, starting from the initial state and initializing inputs
%% \texttt{enable} and \texttt{secret} with symbolic values $\alpha$, and $\beta$,
%% respectively, would yield two leaf nodes: $n_1 = (\sigma
%%
%% , one in which \texttt{enable} is high
%% and one in which it is low. a tree with 8 leaf nodes corresponding to 8 paths
%% through the design. Let \texttt{enable} and \texttt{secret} be initialized with
%% symbolic values $\alpha$ and $\beta$. Following the one-clock-cycle path from
%% the initial state in which \texttt{enable} is high would end 
%%
%% the 4-clock cycle path in Figure~\ref{fig:toy1tree} in which information flows
%% from \texttt{secret} to \texttt{led} has path constraint $\pi = ()$ and ends in symbolic
%% state $\sigma = ()$. 
%\cks{Kaki, please Double check that I'm not
 %off by one.} 

A single path through the IF graph can correspond to many paths
through the symbolic execution tree. For example, \texttt{enable} can remain low
for $0, 1, 2, \ldots$ clock cycles between each update to \texttt{state}. Each
of these options represents a separate path through the symbolic execution
tree. This example, although simple, is not all that contrived. It may be that
state in another module of the design can take
varying time to compute an action before \texttt{enable} becomes high again.
%% \cks{and yay! static analysis enables the unsat core
  %% heuristic and the segmenting by clock cycles, addressing both problems. Make
  %% sure that is made clear later when we refer back to these problems}:
Two problems become apparent:
\begin{enumerate}[noitemsep]
  \item The choice of path in the current clock cycle can determine
    whether there exists a path in a future clock cycle that will allow the flow
    of information to continue.
  \item Once exploration starts down one path, it is not clear at what point -- after how many
    clock cycles -- the current path should be abandoned as incorrect, and a new
    path should be tried. For example, there are infinitely long paths in which
    \texttt{prev} never gets to 3 and \texttt{enable} remains low.
\end{enumerate}

\section{Methodology: Symbolic Execution for Information Flow}
\label{sec:design}
Given a design and an input signal of interest, \texttt{src}, our goal is to
find how information can flow during execution from \texttt{src} through the design. Our approach
is to first use the IF graph to enumerate all potential paths of information flow
through the design from \texttt{src}. As this is a static analysis, complexity
grows linearly with the number of variables in the design and the length of the
RTL code. Then, for each enumerated path, SEIF uses symbolic execution to
either find a corresponding information-flow path through the design, or
determine that no such path exists.

\subsection{Overview}

Once the IF graph is generated, the analysis proceeds in three main phases: pruning globally unrealizable paths,
symbolically executing the design to find realizable paths through the design,
and analyzing the semantics of each found path to find true paths of information
flow. In the following sections, we describe each phase in more detail.

If SEIF returns a path, it is a true path through
the design corresponding to the path in the IF graph. Depending on the
post-processing option used, this will either be a path starting at the design's
reset state or an intermediate state. 

If SEIF does not return a path, there are three
possibilities. First, the path in the IF graph has been identified
as infeasible within a bounded number of clock cycles.
Second, the path in the IF graph is
feasible in the design, but does not represent an actual flow of information --
this result is sound with one caveat discussed in
Section~\ref{sec:semantic}. Third, the path in the IF graph cannot be
accounted for. These options are discussed in
Section~\ref{sec:postprocessing} and evaluated in Section~\ref{sec:eval-acctingpaths}.

%The IF graph is used in the first two phases. In the first phase, information taken directly from the IF graph
%can be used to eliminate paths in the IF graph that are unrealizable. These are
%paths that would require mutually contradictory constraints to hold in order to
%be realized. These are globally unrealizable paths because the constraints are
%unsatisfiable regardless of the current state of the design during execution.

%In the second
%phase, we use symbolic execution to find, for each remaining path through the IF
%graph, a corresponding path through the design. Our methodology relies on the hardware-oriented symbolic 
%execution engine described in \cite{ryan2023countering}. The number of possible paths
%through the design is large and paths are of unbounded length. The IF graph enables pruning some paths at each clock cycle
%boundary and enables a search heuristic to guide execution toward paths
%more likely to complete the flow of information from \texttt{src}.

%In the third
%phase, we use the symbolic state and path condition from the symbolic execution
%engine to eliminate paths that represent true paths through
%the design, but which do not correspond to a true information flow from \texttt{src}.

\subsection{Pruning Globally Unrealizable Paths from the IF Graph}
\label{sec:pruningglobally}

In the first phase, our goal is to quickly and cheaply eliminate paths through the IF graph that are easily
falsified before moving on to the next, more expensive phase. Consider
the example code in Figure~\ref{fig:toy2}. The variable \texttt{temp} carries the input
\texttt{secret} only when the input signal \texttt{enable} is high. The \texttt{secret}
information is conditionally passed on to \texttt{result} and from there to
\texttt{led2}. The corresponding IF graph is shown in
Figure~\ref{fig:toy2fig}. While the IF graph appears
to show a flow of information from \texttt{secret} to \texttt{led2} via
\texttt{temp}, the constraint for edges (\texttt{secret},
\texttt{temp}) and (\texttt{temp}, \texttt{result}) require \texttt{enable} to be high and low, respectively. Since both edges must occur in the same clock cycle,
this flow cannot be realized.

   \begin{figure}[h!]
     \centering
     \begin{framed}
     \begin{lstlisting}[style={verilog-style},belowskip=-.8\baselineskip,aboveskip=-.5\baselineskip]
wire temp = (enable) ? secret : 0;
    
always @(posedge clk) begin
   if (enable) begin
      result <= 0;
   end else begin
      result <= temp;
   end
end

assign led2 = result;
     \end{lstlisting}
   \end{framed}
  \caption{A toy example illustrating globally unrealizable paths. \texttt{clk},
    \texttt{enable}, and \texttt{secret} are input wires and \texttt{led2} is an
    output wire. \texttt{result} is a state-holding reg. \texttt{secret} cannot flow through \texttt{temp} to
    \texttt{result} and \texttt{led2}.}
  \label{fig:toy2}
  \end{figure}

\begin{figure}[h]
  \centering
  \setlength{\belowcaptionskip}{-5pt}
  \includegraphics[scale=.25, trim = -200 150 0 50, clip]{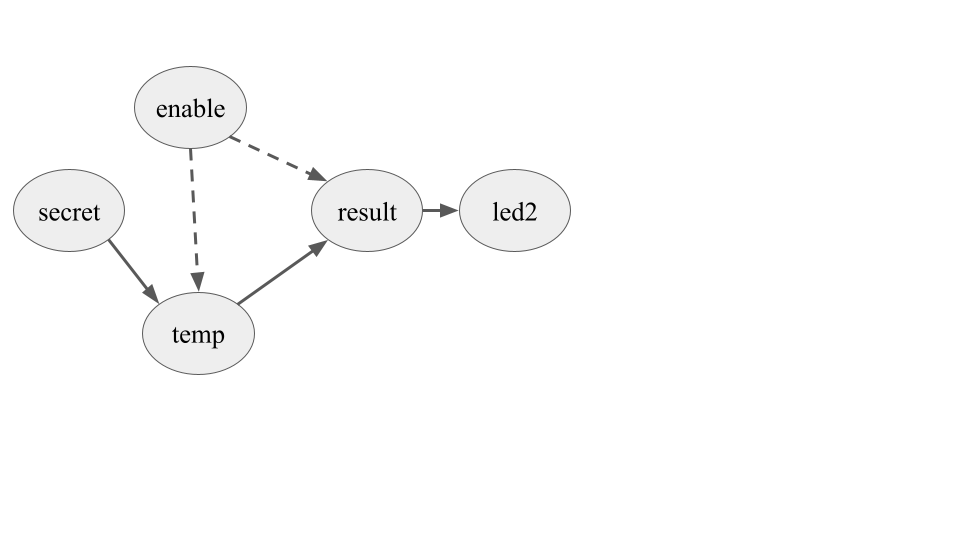}
  \caption{The partial IF graph for the code shown in~\ref{fig:toy2}, showing
    only the paths through \texttt{temp}. Although the graph shows a path from
    \texttt{secret} to \texttt{led2}, an SMT query finds
    that the constraints along the path will never be co-satisfiable.}
  \label{fig:toy2fig}
\end{figure}

This analysis requires knowing where clock
cycle boundaries are. In the IF graph, an edge corresponding to a nonblocking
assignment (for example, $\mathtt{result} <=
\mathtt{temp}$) denotes a clock cycle boundary. When \texttt{state} is updated in
one clock cycle, the updated value can be read in the next clock cycle.

At the start of this phase, the given path through the IF is divided into
\emph{segments}.
One segment of an IF-graph path is a sequence of hops in the IF graph.
These hops could be any implicit or explicit flows. However, the explicit non-blocking assignments are of particular interest to us
in determining how we should break the IF path into segments. Each non-blocking assignment
represents exactly where we reach a clock cycle boundary in the IF path and thus
break off a new segment after that flow.
If a path
has $n$ nonblocking assignments, it has $n+1$ segments.
Let us take the following IF path in Figure \ref{fig:toy2fig} as an example:
$\langle (\mathtt{secret}, \mathtt{temp}), (\mathtt{temp},\mathtt{result}),
(\mathtt{result}, \mathtt{led2})\rangle$.
This path has two segments. The first segment is the two-hop sequence,
$\langle (\mathtt{secret}, \mathtt{temp}), (\mathtt{temp},
\mathtt{result})\rangle$, made up of a continuous assignment and a non-blocking assignment.
The second segment, $\langle (\mathtt{result}, \mathtt{led2})\rangle$, is a single hop and a continuous assignment.

For every segment in a given IF-graph path, the conditions
involved in that segment are collected and checked for co-satisfiability. If
the hops in any one segment have mutually contradictory constraints,
that path is discarded. In Figure~\ref{fig:toy2fig}, the segment $\langle
(\mathtt{secret}, \mathtt{temp}), (\mathtt{temp}, \mathtt{result})\rangle$ has
contradictory constraints, as the first hop requires that \texttt{enable} is high, while the second hop requires it to be low.

This pruning analysis is sound---only unrealizable paths are discarded---as long as the co-satisfiability check considers only state-holding signals and input signals in the
satisfiability query, as these signals do not change value in the middle of
a clock cycle.

\subsection{Symbolic Execution to Find Paths through the Design}

In the second phase, the goal is to find true paths through the design for each
remaining path in the IF graph. We use symbolic execution to find
a sequence of
machine states and a corresponding sequence of input signals (for example, as
seen in Figure~\ref{fig:toy1states}) that aligns with the
path outlined by the IF-graph path.

\subsubsection{Symbolic Execution Guided by IF-Graph Path Segments}
The segment analysis done in the first phase provides information about where
the clock cycle boundaries lie; the IF graph also provides information about
which lines of code must execute for each hop in a segment. SEIF uses
this information to drive symbolic execution along the path outlined by the IF graph.

In each clock cycle, the symbolic execution engine is restricted to following only
those design paths which include the lines of code that must be executed for the current
IF-path segment to be realized. For example, in Figure~\ref{fig:toy1}, the
symbolic execution engine only considers paths which take the \texttt{if}
branch at line 8, when $\mathtt{state == 3}$. By doing so, the search space is significantly reduced.

\begin{figure}[h]
  \centering
  \setlength{\belowcaptionskip}{-15pt}
  \includegraphics[scale=.3,trim= 250 275 425 0,clip]{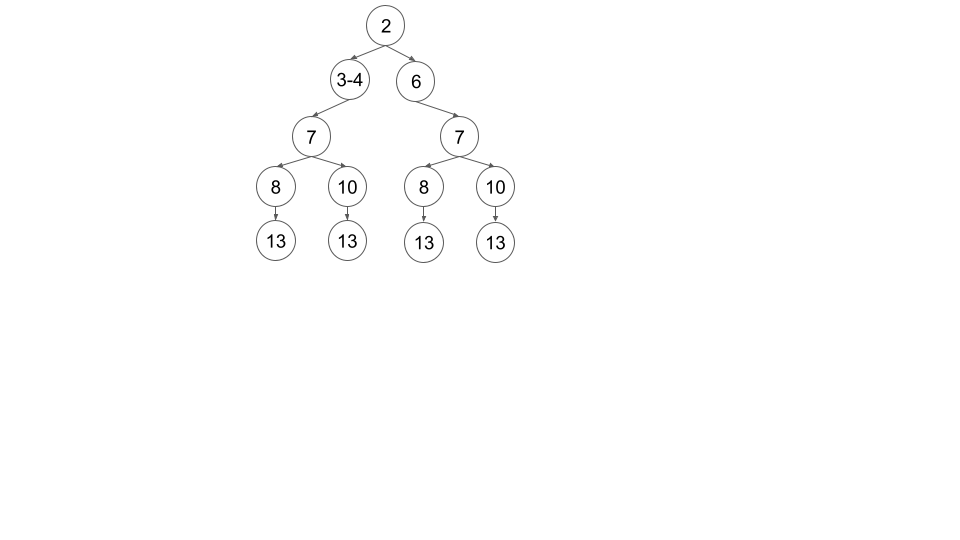}
  \caption{Symbolic Execution Tree of Paths}
  \label{fig:se-tree-paths}
\end{figure}

However, there may still be many possible paths through the design to consider,
only some of which allow the complete IF path to be realized. Continuing
with our example, Figure~\ref{fig:se-tree-paths} shows the symbolic execution
tree for one clock cycle of the code in
Figure~\ref{fig:toy1}. Each node in the tree represents a line of code, or non-branching sequence of code (e.g., lines 3-4) to be
executed.

The path of interest, this time annotated with which line of code needs to
execute for each hop to be realized, is $\langle
(\mathtt{secret}, \mathtt{guard})_{\mathrm{line~} 8}, (\mathtt{guard},
\mathtt{led})_{\mathrm{line~} 13}\rangle$. Examining the symbolic execution tree
in Figure~\ref{fig:se-tree-paths}, it would appear that two of the four
possible paths achieve the desired flow. But annotations
in the IF graph tell us that the
sequence of conditions $(\mathtt{state} == 3)_{s3}, (\mathtt{prev} == 3)_{s4}$
needs to be met. For that to happen, lines 3-4 need to
execute in the first four clock cycles and lines 8, 13 need to execute in only
the fourth clock cycle. While this is clear to see when examining the
state transition diagram (Figure~\ref{fig:toy1states}), there is nothing in the IF graph, or even the
code itself, indicating that it will take four clock cycles to realize this flow. Finding the desired path through the multi-clock-cycle symbolic execution tree is a search problem. We discuss the search strategies we developed to guide search in SEIF in Section \ref{sec:strategies}

%%  Thus, even after forcing the appropriate conditions in each clock cycle, the
%%  resulting tree of paths through the design might still be quite
%%  large. The number of paths to explore is still large. Some of those paths will be infeasible. Some of
%% those paths are feasible, but advance the design to a new state in which
%% down-path segments are unrealizable. Sometimes the design must be advanced to a
%% new state before the next segment can be realized. We discuss each of these
%% conditions, and how they are handled, in the next sections.

%% \cks{Can each subsection be matched to a problem? E.g., infeasible paths in the
%%   SE tree,
%%   advancing to a state which will make down-path segments unrealizable,
%%   advancing to a state required in order to make down-path segments realizable,
%%   large search space}

\subsubsection{Pruning Unrealizable Paths at Clock Cycle Boundaries}
As a first strategy, the symbolic execution engine prunes unrealizable
paths at each clock cycle boundary. At each clock cycle, the engine first
checks the co-satisfiability of the conditions required in the current IF
segment, similar to the check done to prune globally unrealizable paths
(Section~\ref{sec:pruningglobally}). However, this time the SMT query
includes the current symbolic state along with the conditions required for the
IF segment. As with the global pruning step, the check considers only the
state-holding variables in the segment conditions, as the value of combinational
logic variables may change during the course of a clock cycle.

Continuing with
our example from Figure~\ref{fig:toy1}, at the start of the initial clock cycle,
the symbolic execution engine checks whether the condition required for the
first hop in the IF graph ($\mathtt{state} == 3$) is
mutually contradictory with the initial symbolic state (in which $\mathtt{state}
== 0$). Indeed, it is, and the symbolic execution engine discards any paths
that would include line 8, the line of code required for the first hop in the IF
graph.\footnote{Discarding these paths can be done prior to exploration of any paths in
the current clock cycle, as the engine has information from the design's
statically built control
flow graph about which lines of code
are included in which path.} At this point, SEIF recognizes that realizing
the first segment of the IF graph at the current state (state $s0$) is infeasible.

\subsubsection{Stalling the IF-Graph Path to Advance to a New Machine State} 
\label{sec:stalling}

The second strategy used by SEIF is to pause the search for realizing a
segment of the IF path in order to advance the design to a next-state
when needed. In our example, the first segment of the IF graph cannot be
realized from the initial reset state. SEIF symbolically executes the
design for a single clock cycle, without considering the constraints required by
the next IF path segment, to advance the design to a new state. SEIF then
checks whether the IF graph segment can be realized from this new state.

There are many possible next states and SEIF must find one that satisfies two
criteria:
\begin{enumerate}
  \item The next state advances the design toward a state in which the next IF
    segment can be realized, and
  \item The next state does not undo any prior progress along the IF graph path
    that has already been made.
\end{enumerate}

We discuss search strategies for
finding valuable next-states in the next section.
The second constraint is trickier. During normal execution, it is likely that information written to a \texttt{reg} in one clock cycle gets
overwritten in a subsequent clock cycle. For example, consider the code in
Figure~\ref{fig:toystalling}, which is similar to that of our first example
(Figure~\ref{fig:toy1}), but made slightly more complex by the addition of two
new registers: $\mathtt{guard0}$ and $\mathtt{clear}$.

\begin{figure}[h!]
  \centering
    \setlength{\belowcaptionskip}{-10pt}
  \begin{framed}
        \begin{lstlisting}[style={verilog-style},belowskip=-.8\baselineskip,aboveskip=-.5\baselineskip]
always @(posedge clk) begin
   if (enable) begin
      prev <= state;
      state <= state + 1;
   end

   if (state == 0) begin
      guard0 <= secret;
   end else if (clear) begin
      guard0 <= 0;
   end else begin
      guard0 <= guard0;
   end 
          
   if (state == 3) begin
      guard <= guard0;
   end else begin
      guard <= 0;
   end 
   
end
assign led = (prev == 3) ? guard : 0;
    \end{lstlisting}
  \end{framed}
  \caption{A design demonstrating the challenges of stalling.}
  \label{fig:toystalling}
\end{figure}

The IF path of interest is now from \texttt{secret} to \texttt{guard0} to
\texttt{guard} to \texttt{led}. To achieve the second flow segment,
$\langle(\mathtt{guard0},\mathtt{guard})\rangle$, SEIF needs to first
advance the design to a state $s' = \langle\mathtt{state} == 3\rangle$. However, it is
important that while the design advances to state $s'$, the \texttt{clear}
signal is never set, as a 0 written to \texttt{guard0} would undo the information flow from \texttt{secret} to
\texttt{guard0} from the prior IF path segment.

SEIF uses information from the IF graph to \emph{stall}
the information flow while advancing the design to a next-state. We define
stalling as symbolically executing the design for a single clock cycle, such that
the design transitions to a next state, but the position along the IF path
remains unchanged. To stall, SEIF prevents the symbolic execution engine from considering any paths of
execution that will undo information flow from prior segments in the IF
path. To do this, SEIF considers the node $n$ in the IF path, in which information
currently ``resides.'' In our current example, this would be the node
\texttt{guard0}. SEIF then uses the IF graph to find all edges incident to node
$n$, which represent flows of information from variables in the design to $n$ and are associated with lines of code. Explicit
flows need to be prevented during stalling, but implicit
flows do not need to be prevented, as they do not cause the value in $n$ to be overwritten. SEIF avoids exploration of any paths through
the design which would execute a line of code in which $n$ is written to. In
this way, the information in $n$ is not lost while stalling. 

There are two edge cases to consider. The first is self-loops. Direct flows from $n$ to $n$ (e.g.,
$\mathtt{n} <= \mathtt{n} + 1$ are
allowed, as the information in $n$ stays in $n$. The second is the case when $n$
is assigned a constant (e.g., $\mathtt{n} <= 1$). SEIF
checks this corner case during symbolic execution and abandons any path in
which it occurs. If the assignment by a constant happens regardless of the rest of the state,
then stalling cannot occur at this point in the IF graph.

Because of stalling, the number of clock cycles needed to verify the information flow
may exceed the length of the IF path.

\subsubsection{Search Strategies}
\label{sec:strategies}
The goal is to find a sequence of design states, and corresponding
input values, that correspond to an IF path, or determine that
no such sequence exists. The search space is large; an IF path with $n$ segments
requires at least $n$ clock cycles through the design. When stalling is
needed, the number of clock cycles required is 
unbounded (although finite).

Information from the IF graph is used to prune the symbolic execution tree at each clock cycle, but a single IF hop can correspond to many paths through the symbolic
execution tree. This is because a segment of the IF-path involves only a small number (typically
fewer than 5) of lines of code be executed. The input space is partially
constrained to ensure
those few lines of code are executed, but most of the input space is unconstrained,
and therefore there is freedom in how most of the design is explored at each
clock cycle.

%And a poorly chosen path through the design in an early clock
%cycle may preclude successfully finding a needed path through the design in a
%later clock cycle.
 
  %\kar{cut this?} Additionally, at each clock cycle it may be the case that information flow can stall
    %for an arbitrary or fixed amount of time. In other words, the IF-path can
    %stutter while the design continues to execute for some number of clock
    %cycles, and then the IF-path is picked back up again.
 
We developed and implemented four search strategies:
\begin{itemize} 
  \item Baseline 1 --  Continue / Stall Only.
  \item Baseline 2 -- Backtracking Only
  \item Stalling with Backtracking
  \item Stalling with UNSAT Core Heuristic
  \end{itemize}
  
 \paragraph{\textbf{Baseline 1: Continue / Stall Only}}
 \label{sec:baseline-1}
The key idea behind this strategy is that, for each segment, we can either symbolically
execute until a design path is found in which the segment conditions are 
satisfied (termed a \emph{continue}), or we can stall for some bounded number of cycles. For an IF path, we build and exhaustively search a list of all possible \emph{continue}, \emph{stall}
combinations. If SEIF is unable to complete the IF path for a given
continue-stall pattern, it moves on to the
next pattern. The list of continue-stall combinations are in truth-table order to allow the SEIF
engine to explore as deeply as possible first, aiming to verify the shortest path possible with no stalls.
In this context, depth equates to the number of IF-path segments successfully
traversed, and for which SEIF has realized a partial path of execution.
      %then there are two options: stall or backtrack. Use heuristics \cks{from the unsat core we think}to determine which option to chose.
    %\item If stalling, choose one of CCSC, CSCC, etc. and either start the
      %search fresh, this time with a stall cycle in between two continue cycles,
      %or go back to an earlier clock cycle in this search and add a stall
      %cycle. Our strategy proceeds in truth-table order. \kar{need to explain better probably}
      %\item With this truth-table ordering, we go deep as possible first. The idea being that we may as well try for
      %the shortest possible path with no stalls first. In this context, we mean depth to mean the number of information
      % flow segments successfully traversed, and for which we have a path through the symbolic execution tree. We have
      % found realizable paths through the design for these partial flows. Breadth in this context through a symbolic execution
      % tree would be exploring all of the paths until we find one that produces the necessary flow for a particular information
      % flow segment.
      %\item Once all SE paths for this cc have been explored and none
      %    have produced the information flow that we desire for this IF-path
       %   segment, then we know we need to either a) find a different SE path in
       %   one of the earlier clock cycles or b) stall somewhere along the
       %   IF-path -- maybe here or maybe earlier. The order we proceed in is defined for us -- truth-table.

 \paragraph{\textbf{Baseline 2: Backtracking Only}}
In this search strategy SEIF
begins by symbolically executing for one clock cycle for the first segment. If the flow found, SEIF 
moves to the next segment in the IF path. If at any segment, the flow is not found in some bounded number
of clock cycles, or there are no more design paths to try, SEIF returns (or \emph{backtracks}) to an earlier
segment to find a different path that satisfies the same segment conditions.

%Some bookkeeping is done here to be able to re-search the second clock cycle to look for a new path
%without repeating work, too. \kar{explain}

% \begin{itemize}
 %\item  Go back to an earlier clock cycle to look for a different path that satisfies
%the same IF-path-segment flow requirement. 
 %\item If backtracking, go back to the prior clock cycle and look for a
 %     second path in which the leaf node shows that information has flown from
 %     $x$ to $y$ -- again, not looking at leaf nodes, just whether or not a line
 %     of code has been executed. Some bookkeeping is needed to be able to re-search the second
 %     clock cycle to look for a new path without repeating work.
% \end{itemize}
 
  \paragraph{\textbf{Stalling with Backtracking}}
   This strategy is a hybrid of baselines 1 and 2. 
   For any given \emph{continue}, \emph{stall} pattern, after successfully
   executing consecutive \emph{continues}, and reaching a \emph{stall}, SEIF
   stalls for a bounded number of clock cycles and
   attempts to find execution paths where SEIF can make forward progress in the next segments. If all symbolic execution paths are explored, or SEIF times out (according to some pre-determined bound), it backtracks.

\paragraph{\textbf{Stalling with Heuristic}}
    \label{sec:unsat}
   This strategy builds on top of stalling with backtracking. Our heuristic
   relies on the \emph{UNSAT core}, the subset of constraints in a SAT query for
   which no satisfying assignment exists. If SEIF stalls, it is searching for a
   new machine state that will satisfy the conditions of the next IF path segment. 
In this case, SEIF pushes the symbolic state and the constraints from the next 
   segment to the SMT solver, which returns the UNSAT core. %%  From these constraints we derive the UNSAT core, which helps to find a path in the design that (1) corresponds to a stall of the information flow, and (2) makes
   %% forward progress along the IF flow path. To do this, after stalling,
   For each path explored while stalling, SEIF checks if the
   UNSAT core became smaller. If it did, SEIF continues searching for a new
   machine state along the path. If it grows, SEIF prioritize the next candidate stall path.

   \subsubsection{Post Processing to Find Reset}
   \label{sec:postprocessing}
       SEIF begins exploration from a symbolic state, and therefore the
       design paths it generates inputs for may not start from the
       reset state. We mitigate this by checking whether the found
       design path has constraints that conflict with the design's reset
       state. If not, the path can start from reset. If so, the path
       starts from an intermediate state of the design, and SEIF cannot
       guarantee that it is a reachable state. Most often, SEIF finds paths that
       can start from reset and we evaluate this in Section~\ref{sec:eval-acctingpaths}.

\subsection{Semantic Analysis to Identify True Information Flows}
\label{sec:semantic}
Once per execution path, SEIF performs a semantic analysis check to prune 
flows that represent viable design paths, but not true flows of information. 
This can happen when a textual flow does not represent an information flow. For example, \texttt{y <= x  xor  x}, would yield a path showing x flows
to y even though there is no flow from x to y. SEIF prunes explicit textual
flows which do not represent information flows.

If there is an implicit textual flow that is not a true information flow, SEIF cannot eliminate
that false positive. For example, \texttt{if (x XOR x) y <= 0; else  y <= 1;}
 (Here, there is no path in which \texttt{y} is set to 0, and SEIF does recognize
that.)
    
%% \begin{figure}[h]
%%   \centering
%% \begin{subfigure}[b]{0.2\textwidth}
%% %   \begin{framed}
%%      \begin{lstlisting}% [linewidth=.2pt,style={verilog-style},belowskip=-.8\baselineskip,aboveskip=-.5\baselineskip]
%% x
%%      \end{lstlisting}
%%  %  \end{framed}
%%   \caption{Implicit textual flow}%An implicit textual flow that is not a true flow of information.}
%%   \label{fig:text-flow}
%%   \end{subfigure}
%%  \begin{subfigure}[b]{0.2\textwidth}
%% %   \setlength{\belowcaptionskip}{-10pt}
%%   \includegraphics[scale = 0.3, trim = 30 0 40 10, clip]{figures/reconvergent.png}
%%   \caption{Reconvergent flows}
%%   \label{fig:reconvergent}
%%  \end{subfigure}
%%  \caption{Textual information flows}
%%  \label{fig:semantic-if-flows}
%% \end{figure}
%% \vspace{-10pt}

\begin{figure}[h]
  \centering
  \includegraphics[scale = 0.3, trim = 30 0 40 10, clip]{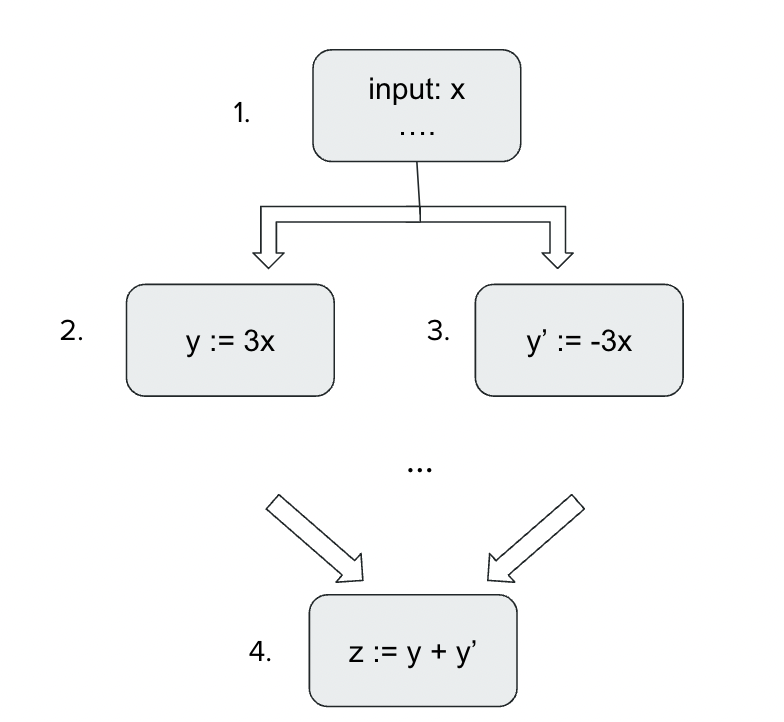}
  \caption{Reconvergent flows}
  \label{fig:reconvergent}
  \end{figure}

In the case of reconvergent fan-out SEIF may or may not find the flow. In
the example of Figure \ref{fig:reconvergent}, \texttt{x} is an input and blocks 2 and
3 represent different areas of the design (i.e. modules, always blocks). There are 4
cases to consider: 
\begin{enumerate}
\item The writes to \texttt{y} and \texttt{y'} are both unconditional and there is no flow from \texttt{x} to \texttt{z} because $\mathtt{z} = 3\mathtt{x} - 3\mathtt{x}$.
SEIF performs the check and correctly detects no flow.
\item The writes to \texttt{y} and \texttt{y'} are conditional, and depend on
  the same conditions. SEIF detects there is no flow
\item The write to \texttt{y'} is conditioned on something that is mutually UNSAT with the condition for \texttt{y}. In this case, there 
is always a flow from \texttt{x} into \texttt{z}, and SEIF detects it.
\item The write to \texttt{y'} is conditioned on something mutually satisfiable with the condition for \texttt{y}, where the condition
for \texttt{y} is different. If SEIF follows a design path where both conditions are true at the same time, it detects no flow,
while there may be other design paths through block 3 which would enable a flow,
and vice-versa. Unless SEIF is able
to exhaustively explore, it may report an incorrect result.
\end{enumerate}
\vspace{-5pt}
    \vspace{-10pt}

%In the case of an implicit flow at the textual/signal level from $x$ to $y$, but
%where there is no flow at the semantic level, we do not differentiate.

\section{Implementation}

We implemented SEIF using the Sylvia symbolic execution
engine~\cite{ryan2023countering} and using hyperflow
graphs~\cite{meza2023hyperflowgraph} as our IF graph engine.\footnote{We
contacted the authors of the hyperflow graph paper and they gave us closed-box
access to the tool, providing the static analysis for the designs we gave
them. The
authors of the Sylvia symbolic execution engine gave us source-code access to
their tool.} Both Sylvia and
the hyperflow graph toolchain were built using python3. Sylvia
implements the Verilog semantics according to the 
IEEE 1364-2005 semantics using pyVerilog and the Z3 solver for SMT solving. SEIF
also uses Z3 for preprocessing and path removal.

When considering information flow paths that span multiple modules, enumerating all possible 
paths for even a single source/sink pair becomes too
expensive. We manage this complexity by following the divide-and-conquer
approach of Ryan et al.~\cite{ryan2023countering}.
SEIF first finds the partial IF paths within a module, and then uses the segment
conditions to find the next module to explore. SEIF uses the SMT solver to ensure that the path fragments can be stitched back together to form a valid information flow path from source to sink. 
This approach reduces repeated work within
a module when exploring paths across multiple modules.

\section{Evaluation}
\label{sec:eval}

We evaluate SEIF over four open-source designs to study its viability 
as a means for accounting for information flows within a hardware design. The evaluation
addresses the following questions: 1) Can SEIF recognize and eliminate paths
through the IF graph that are unrealizeable in practice? 
2) Can SEIF find paths through the design, 
along with the sequence of inputs to realize the path, that corresponds to paths through the IF graph?
3) Can SEIF be meaningfully applied to security relevant signals in hardware
designs to give experts 
feedback on the security of the design or new areas to explore?

\subsection{Dataset and Experimental Setup}

We collected four open-source Verilog designs for evaluation. The designs are:
\begin{enumerate}
	\item OR1200, a 5-stage RISC processor core~\cite{or1200};
	\item openMSP430, a synthesizable 16-bit microcontroller core~\cite{msp430};
	\item the AKER Acess Control Wrapper (ACW)  \cite{restuccia2021aker}; and
	\item an AES implementation from TrustHub \cite{TrustHub1} \cite{TrustHub2}.
\end{enumerate}

The experiments are performed on a machine with an Intel Xeon E5-2620 V3 12-core
CPU (2.40GHz, a dual-socket server) and 62G of available RAM. 

%% In the following sections we For each design we provide some performance metrics, statistics to answer the research questions above, and security analysis
%% focused on the specific security critical signals or properties of interest.

\subsection{Accounting for Paths in the IF Graph}
\label{sec:eval-acctingpaths}
We first examine SEIF's ability to account for paths in the IF graph,
either by finding paths through the design that correspond to the IF path, or by
eliminating the IF path as infeasible. In these experiments we look at the OR1200, MPS430, and ACW
designs, which are the largest of the four.

We identified 20 security-critical signals in the OR1200 to use in our experiments that appear in
security properties of the OR1200 collected from the security literature \cite{hicks2015specs} \cite{bilzor2011security} \cite{zhang2017scifinder} \cite{zhang2020transys} \cite{trippel2020ICAS}. We selected 10 sources to analyze in the MSP430 by finding signals roughly analogous to those in the security properties for the OR1200.   
For the ACW, we chose 20 main internal signals to look at that appear in the security properties manually and automatically generated by \cite{restuccia2022framework} \cite{Deutschbein2022JCEN} and map to several known CWEs. 

For each source signal there can be tens of thousands of IF paths. (See the numbers in Tables~\ref{lst:security-analysis-msp430}
and~\ref{lst:security-analysis}, discussed in Section~\ref{sec:security-analysis}.) For the efficacy
and performance evaluations in this and the next two sections, we analyze a subset
of the total paths. For each source signal of interest, we randomly selected 300
paths from the IF graph for analysis. For the security analysis case study
(Section~\ref{sec:security-analysis}), we analyze all paths from a given source.

%\subsection{Accounting for Information Flows}

\begin{figure}[h]
  \centering
  \setlength{\belowcaptionskip}{-5pt}
    \includegraphics[scale=0.28, trim = 0 20 0 55,clip]{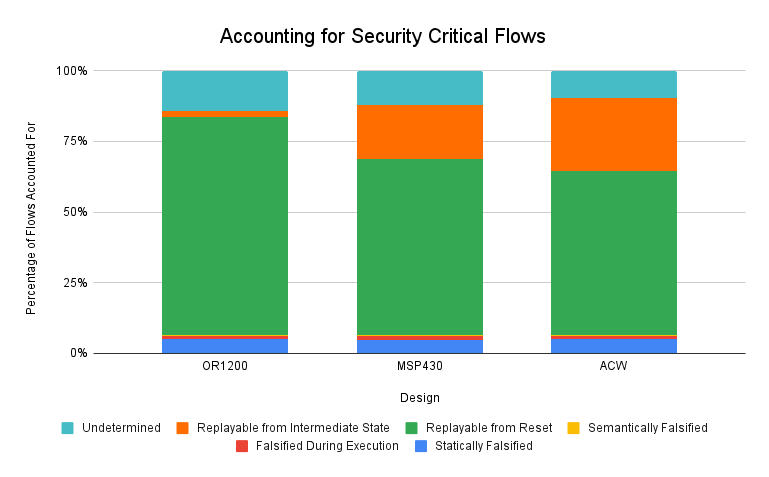}
\caption{Accounting for IF Paths}
    \label{fig:accounting}
\end{figure}

%% For the OR1200 and ACW, we first identify security critical sources, and then we choose 300 random paths to perform the information flow analysis. The numbers in the table
%% represent the average taken over all sources. 
Figure~\ref{fig:accounting} summarizes SEIF's ability to account for the IF
paths.
For 86\% to 90\% IF paths on average, SEIF either finds the corresponding path through the
design or eliminates the IF path as infeasible or not representing a true flow
of information. The majority of accounted-for IF paths, 58\% to 77\% on 
average in the three designs, are true paths in the design, indicating that the
static analysis done to build the IF graph is a decent approximation of
information flow through the design. We further break down these
numbers to show the percentage of the found IF paths for which SEIF returns a
design path that starts at the reset state vs. a design path that starts at some
intermediate state. Paths that start at the reset state are better for the
engineer as they can be immediately replayed from the known reset state.

\subsection{Evaluation of Search Strategies}
\begin{figure}[h]
  \centering
    \includegraphics[scale=0.35,trim = 0 20 0 55,clip]{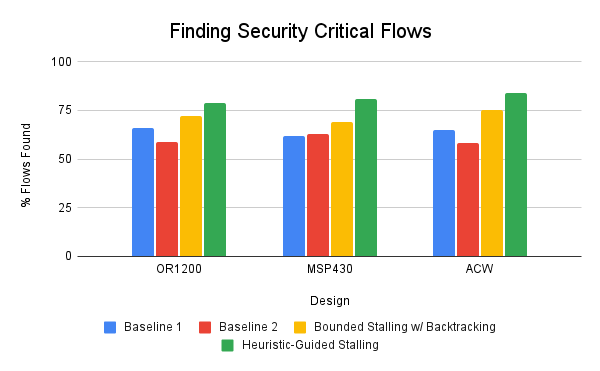}
\caption{Finding Design Paths Corresponding to IF Paths}
    \label{fig:findflows-percent}
\end{figure}

In the following we evaluate the four search strategies discussed in
Section~\ref{sec:strategies}. Figure~\ref{fig:findflows-percent} reports, for
each design, the percentage of IF paths found by each of the four search
strategies. These are paths for which SEIF found a corresponding path through
the design. As expected, the heuristic
guided search outperforms the other strategies in all three designs, improving
over the baselines by 26\% on average and over bounded stalling with
backtracking by 11\% on average. We note that baseline 2, which does
not include stalling, is the least successful at finding corresponding paths in
the design. This highlights the value of SEIF: many IF paths give an incomplete
picture of a path through the design and include points where the design must advance to
a new state before the IF path can continue. Without SEIF, it would be up to the engineers to
figure out how and whether to advance the design state.

\begin{figure}[h]
  \setlength{\belowcaptionskip}{-5pt}
  \centering
    \includegraphics[scale=0.35,trim = 0 20 0 55,clip]{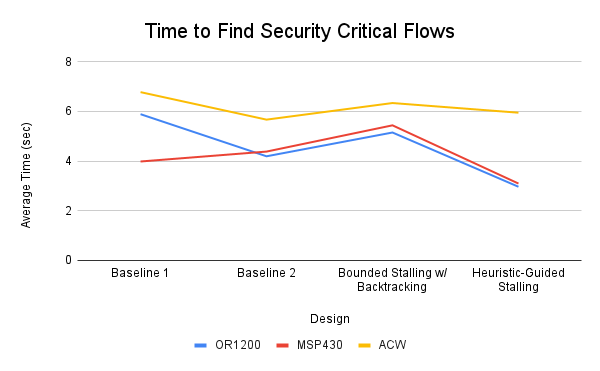}
\caption{Time to Find Design Paths}
    \label{fig:findflows-time}
\end{figure}

\begin{figure}[h]
  \setlength{\belowcaptionskip}{-5pt}
  \centering
    \includegraphics[scale=0.35,trim = 0 20 0 55,clip]{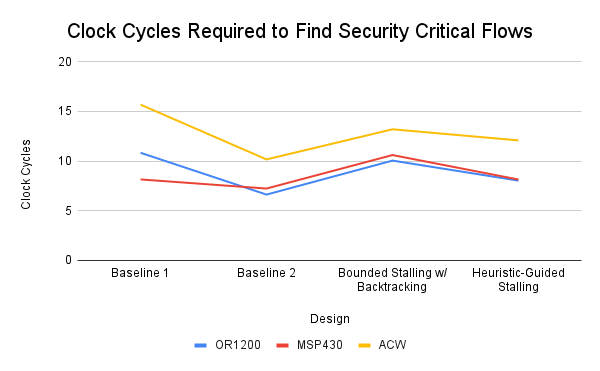}
\caption{Clock Cycles to Find Design Paths}
    \label{fig:findflows-cc}
\end{figure}

Figures~\ref{fig:findflows-time} and \ref{fig:findflows-cc} report on the
performance of the four search strategies, both in terms of average time taken to
find a corresponding design path and average number of clock
cycles through the design for the found path. Again, the
heuristic guided search outperforms the other strategies, completing the search
for each IF path in 3-6 seconds.

\begin{figure}[h]
  \setlength{\belowcaptionskip}{-10pt}
  \centering
    \includegraphics[scale=0.35,trim = 0 20 0 55,clip]{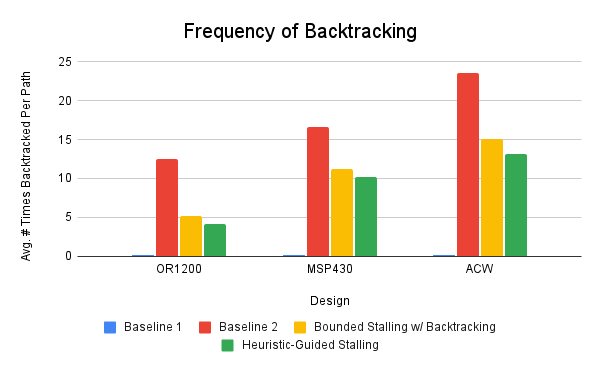}
\caption{Frequency of Backtracking}
    \label{fig:backtracking-freq}
\end{figure}

Figure~\ref{fig:backtracking-freq} shows that the amount of backtracking that is
required is lowered when we incorporate bounded stalling. Adding the heuristic
improves the efficacy of stalling and therefore decreases backtracking even further.

%% how well SEIF performs with the four  heuristic-guided search strategy 
%% Figures \ref{fig:findflows-percent}, \ref{fig:findflows-time},
%% \ref{fig:findflows-cc}, \ref{fig:backtracking-freq} and  \ref{fig:accounting}
%% summarize our the methodology's ability to find information flows. For 300
%% randomly selected paths for each of the security critical sources, we
%% attempted to verify the information flow path. We report the average \% of
%% flows found, clock cycles necessary to find the flow and time taken. This was
%% repeated for each of the 4 search strategies/heuristics.

\begin{figure}[h]
  \setlength{\belowcaptionskip}{-10pt}
  \centering
    \includegraphics[scale=0.35,trim = 0 20 0 65,clip]{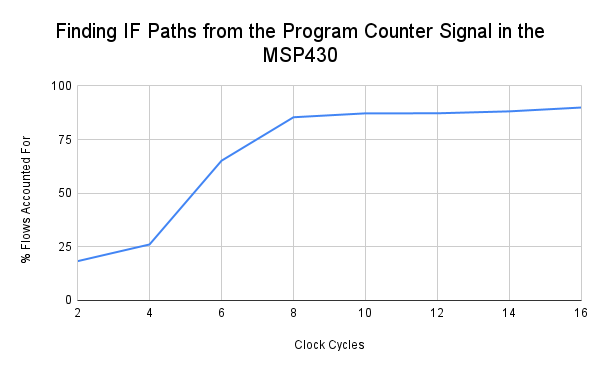}
\caption{Finding design paths over time}%corresponding to IF paths from the program counter in the MSP430 over time}
    \label{fig:flows-overtime}
\end{figure}

To better understand how SEIF is finding flows over time, we explore all
IF paths from a single source signal, the program counter, in
the MSP430. We track how many IF paths are found in the design after 1 clock
cycle of search, 2 clock cycles of search, etc. The experiment was done with
heuristic-guided stalling turned on. Figure~\ref{fig:flows-overtime}
shows the results. There were a total of 19060 IF paths, and SEIF
found design paths for 89.93\% of them. The complete search took 16 clock
cycles, however, most of the paths were found withing the first 8 clock
cycles. The experiment took 3.5 days to run.

  \subsubsection{Determining the Stall Bound} 
For all the experiments in the previous sections, the number of stalls per IF
path segment was set to be 
5, 5, and 4 for baseline 1, bounded stalling with backtracking and the UNSAT core heuristic, respectively. 
(As a reminder, baseline 2 is backtracking only, with no stalling).
We determined these numbers empirically by selecting at random 5 of the
security-critical source signals from the OR1200, and for each of these source
signals selecting at random 300 paths to evaluate, and then running the experiments with an
increasing number of stalls allowed until we saw the number of IF paths found
begin to flatten out. Finding the bound for the heuristic-guided stalling
strategy is shown in Figure \ref{fig:stalling-unsat-graph}. The graphs
for the other three search strategies are in the appendix. 

%\begin{figure}[h]
%  \centering
%    \includegraphics[width=\columnwidth,trim = 0 20 0 55,clip]{figures/baseline1-stalling-graph.png}
%\caption{Empirically determining the stall
%  bound for baseline 1.}
%    \label{fig:baseline1-graph}
%\end{figure}

%\begin{figure}[h]
%  \centering
  %  \includegraphics[scale = 0.45]{figures/bd-stalling-graph.png}
%    \includegraphics[width=\columnwidth,trim = 0 20 0 65,clip]{figures/bd-stalling-graph.png}
%\caption{Empirically determining the stall
%  bound for bounded stalling with backtracking.}
%    \label{fig:bd-stalling-graph}
%\end{figure}

\begin{figure}[h]
  \centering
    \setlength{\belowcaptionskip}{-15pt}
    \includegraphics[scale=0.38, trim = 0 20 0 65,clip]{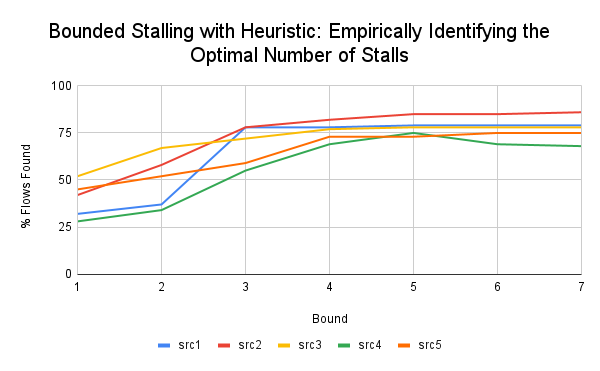}
\caption{Finding the stall
  bound.}
    \label{fig:stalling-unsat-graph}
\end{figure}

\subsection{Eliminating Information Flows Paths}
\begin{figure}[h]
  \setlength{\belowcaptionskip}{-10pt}
  \centering
    \includegraphics[scale=0.38,trim = 0 20 0 55,clip]{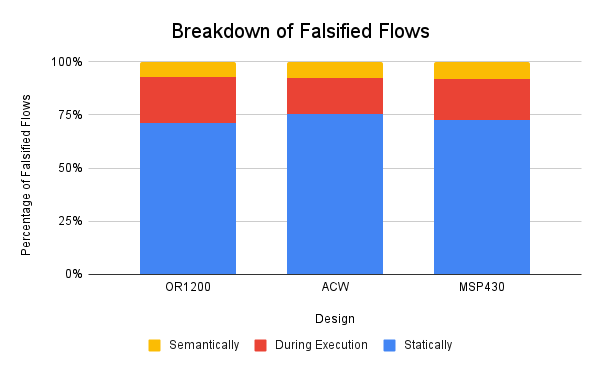}
\caption{Breakdown of how flows are falsified by SEIF}
    \label{fig:falsified}
\end{figure}

We examine how IF paths that do not correspond to information-flow paths
through the design are falsified in Figure~\ref{fig:falsified}. The experiment used the 300
randomly chosen paths for the 20 security-critical signals in the OR1200. The
largest percentage of eliminated paths are found statically before symbolic
execution begins. This is good news, as that is the cheapest and quickest phase
of the analysis. There is a non-trivial portion, 5\% to 7\%, that are eliminated because they do not represent
true flows of information through the design. SEIF's use of symbolic execution
allows for this precise analysis, which taint tracking may not be able to
provide.

%% \cks{Maybe this is totally redundant information with
%%   Figure~\ref{fig:accounting}. Cut?}
%% In Figure \ref{fig:replay}, we report the average number of information flows we still find to be viable after our pruning passes
%% and how many we find to be repayable from the reset state in the OR1200. 

%% \begin{figure}[h]
%%   \centering
%%   %  \includegraphics[scale = 0.45]{figures/bd-stalling-graph.png}
%%     \includegraphics[width=\columnwidth,trim = 0 20 0 55,clip]{figures/replay.png}
%% \caption{Finding replayable information flow paths}
%%     \label{fig:replay}
%% \end{figure}

\subsection{Case Study: Security Property Verification}
\label{sec:security-analysis}
When starting with a property, such as is often done in security verification
tasks, SEIF goes beyond producing a single counterexample. In traditional,
assertion-based formal methods, once the formal or bit-level engine produces the first counterexample, it takes manual manipulation of the property 
or environment to generate subsequent violating traces.
SEIF is able find multiple realizable traces through the design that exhibit the vulnerable behavior and 
can guide the security engineer to other areas of the design they may be interested in exploring. 

We demonstrate the approach for two security-critical properties from the TrustHub
Security Property/Rule Database~\cite{TrustHub1, TrustHub2}, one for the MSP430
and one for an AES implementation. The MSP430 property asserts that the program counter’s value should not be readable
 from the debug access port during normal operation. %% This is a case where information, or a security asset, should 
 %% not be transmitted across certain module boundaries under certain conditions.
The AES property verifies that the secret key material is not accessible to any
 unprivileged internal data registers \cite{zhang2021sidechannel}. 

SEIF generates all the paths from the source of interest to the security-critical sink automatically. 
 In order to produce the violating paths,
 SEIF adds a constraint to the solver specifying the desired precondition. 
 If we find a candidate violation
of the security property, we ensure it is replayable from the reset state of the
design. The results for the MSP430 and AES are presented in
Tables~\ref{lst:security-analysis-msp430} and \ref{lst:security-analysis}, respectively.

%Our methodology would allow a security engineer to quickly enumerate the ports in the frontend module where the program counter originated, and the debug module, and find information flow paths between any of the source/sink pairs, beyond the original two that we identified in the property. 

%We can symbolically execute any one of these paths, and if we successfully find the flow, then we can send the path condition to the solver and it will generate the constraints we need to realize that path concretely. However, this set of paths is a superset of the paths that actually violate the property, during this exploration phase we were not taking into account this precondition -- ``during normal operation.'' The solution is to add an additional constraint to our solver before finding that concrete input trace specifying the desired precondition. 

 \begin{table}[htb]
  \small
  \centering
  \begin{tabular}{rr}
  \toprule
  Metric & Result \\ 
  \midrule
      Total IF paths from source:      & 19060       \\
      Total sinks reachable from source:  & 41   \\
      Total IF paths violating security property: & 58 \\
      Avg. time to produce a counterexample (s): & 0.678 \\ 
      Avg. no of clock cycles explored:  & 8.13 \\
      Total realizable paths violating security property: & 46 \\
      \bottomrule
        \end{tabular}
  \caption{Security Property Verification: Program Counter in MSP430}
  \label{lst:security-analysis-msp430}
  \end{table}

  \begin{table}[htb]
  \small
  \centering
  \begin{tabular}{rr}
  \toprule
  Metric & Result \\ 
  \midrule
      Total IF paths from source:      & 61639       \\
      Total sinks reachable from source:  & 39   \\
      Total IF paths from source violating security property: & 57 \\
      Avg. time to produce a counterexample (s): & 0.505 \\ 
      Avg. no of clock cycles explored:  & 4.102 \\
      Total realizable paths violating security property: & 25 \\
      \bottomrule
        \end{tabular}
  \caption{Security Property Verification: Secret Key in AES Implementation}
  \label{lst:security-analysis}
  \end{table}

\section{Related Work}

\paragraph{\textbf{Symbolic Execution of HW Designs for Information Flow Analysis}}
EISec uses netlist-level
symbolic execution to verify information-flow safety and
quantify confusion and diffusion in cryptographic
modules~\cite{fowze2022eisec}. Our work improves upon EISec by
allowing analysis at the RT-level and enabling verification of a wider class of
information-flow properties. Other tools
use symbolic simulation
(e.g.,~\cite{torlak2014rosette}) %, either in combination with manual
%refinement proofs~\cite{athalye2022knox} or information flow
%tracking~\cite{cherupalli2017software}
to verify particular binaries running on the
hardware~\cite{athalye2022knox,cherupalli2017software}.

%% Work by Cherupalli et al.
%% uses gate-level symbolic simulation in conjunction with software-based information flow tracking flow system to eliminate vulnerabilities in hardware \cite{cherupalli2017software}

%\cite{cherupalli2017software}
%Gate-level symbolic simulation (not SW-style symbolic execution) for finding information flow vulnerabilities in HW/SW -- an IoT processor running the
%particular IoT application. Has comprehensive list
%of references for how information flow analysis (mostly taint tracking, but not
%all) is used to assess security and privacy of systems (mostly
%software). ``symbolic gate-level simulation of the application binary on the
%netlist''. Key information flow policy assessed is memory leakage (tainted to
%untainted output). Evaluation on openMSP430. 

\paragraph{\textbf{Symbolic Execution of SW for Information Flow Analysis}}
%\cks{I think all three use self-composition for SE of IF?}
The software community was perhaps the first to leverage symbolic execution to
verify information flow. The approach has been used in combination with taint
tracking~\cite{cha2012mayhem}, to find and mitigate side
channels~\cite{bao2021symbolic,wang2017cached,wang2019identifying,brotzman2019casym},
and to identify programs that are
vulnerable to transient execution attacks~\cite{guarnier2020spectector}.

%% These techniques have also used in the Mayhem engine, which
%% uses taint tracking to guide execution towards vulnerabilities in binaries~\cite{cha2012mayhem}, as well as in CacheD to discover cache-based timing side channels in production software~\cite{wang2017cached,wang2019identifying}. Spectector~\cite{guarnier2020spectector} uses the symbolic execution to analyze 
%% the flow of information in software. Their goal is to find programs that leak information under 
%% a speculative execution model. Our work carries similar ideas into the hardware domain.
%% CaSym is a cache-aware software symbolic execution engine for side channel detection and mitigation \cite{brotzman2019casym}.

\paragraph{\textbf{Symbolic Execution of SW or HW to Find Exploitable Flaws}}
There is a long history of using symbolic execution in software to find
exploitable security flaws
(e.g,~\cite{avgerinos2011automatic,avgerinos2014automatic,renzelmann2012symdrive}). 
In hardware, symbolic execution has been used to find violations of and exploits
for security-critical assertions~\cite{zhang2018end} and to find and trigger
trojans in the Verilog RTL~\cite{Shen2018SymbolicEB}. As
with SEIF, the main challenge is guiding search through the tree to find the
salient paths.

%% Work on Automatic Exploit Generation~\cite{avgerinos2011automatic,avgerinos2014automatic} has explored preconditioned symbolic execution, which restricts exploration to only likely-exploitable regions of the state space.
%% Our technique shares ideas/philosophy with preconditioned symbolic execution. In
%% PSE, the input space is limited to prune the search tree. In contrast, we add constraints that allow continued information flow, which guides the search through the tree.

%% SymDrive~\cite{renzelmann2012symdrive} is a tool which uses symbolic execution to find bugs in software drivers and allows developers to guide symbolic execution down particular paths.  Coppelia~\cite{zhang2018end} describes a technique for performing backwards symbolic execution to find security violations and generate exploits for hardware. There has also been work using symbolic execution to generate test cases to trigger trojans in Verilog RTL \cite{Shen2018SymbolicEB}.

\paragraph{\textbf{Information Flow Tracking in HW}}
The state of the art for information flow analysis in hardware is information flow
tracking (IFT), which
instruments a design with tracking logic~\cite{tiwari2009complete}. Many tools
operate at the netlist level, although some operate at
the RTL level~\cite{ardeshiricham2017register}. IFT has also been used in analog
designs~\cite{bidmeshki2017information}, and tools exist to synthesize
designs that incorporate 
tracking logic~\cite{pilato2018tainthls,pieper2020dynamic}.
IFT can be used to check
hyperproperties and has been used to verify the safety and security of many different systems~\cite{clarkson2010hyperproperties,Kozyri2022expressing}\cite{tiwari2009execution,boraten2018securing}\cite{meza2022safety}\cite{restuccia2022framework,restuccia2021aker}\cite{ardeshiricham2017clepsydra}
\cite{hu2016detecting}\cite{jin2012proof}. IFT has also been
used to automatically generate information flow properties for use with formal
verification engines~\cite{Deutschbein2021Isadora,Deutschbein2022JCEN}. We used
these properties in our evaluation. 

%and has been used to
%enforce non-interference~\cite{tiwari2009execution,boraten2018securing},
%verify safety of communication protocols~\cite{meza2022safety}, reason about
%access control violations~\cite{restuccia2022framework,restuccia2021aker},
%detect timing flows in hardware~\cite{ardeshiricham2017clepsydra}, find hardware
%trojans~\cite{hu2016detecting}, and to
%ensure data secrecy and hardware trust~\cite{jin2012proof}.

%% \cks{Kaki, any chance we use
%%   any of these properties in our analysis? If so, I'd change the last sentence
%% to say that.}
\paragraph{\textbf{Formal Analysis for Information Flow}}
Proof-checking approaches have been used for detecting
security vulnerabilities in hardware designs \cite{fadiheh2023exhaustive} \cite{kong2017using}.
These approaches are often less automated, more time
intensive, and tackle smaller designs, for stronger results that are both sound and complete.
 VeriCoq translated Verilog to Coq for proof-carrying designs \cite{bidmeshki2015vericoq}. Another approach is
to use self-composition, or program products, to verify information-flow properties~\cite{eilers2021product}.
Security extensions in the hardware description language can enforce 
information flow policies at the language level~\cite{ferraiuolo2017secverilog, deng2017secchisel, li2011caisson, li2014sapper,zhang2015secverilog, ardeshiricham2019verisketch}.

\section{Conclusion}

SEIF combines static analysis and symbolic execution to find
information flows in hardware designs. SEIF improves over static analysis,
eliminating false-positive flows, and finding replayable designs through the
path for true flows. In our experiments, SEIF accounts for 86--90\% of
statically identified flows in three open-source designs. SEIF also leverages
static analysis to explore the designs for 10-12 clock cycles in 4-6 seconds on average.
Additionally, SEIF can be used to find multiple violating paths for security
properties, providing a new angle for security verification.

\section{Acknowledgments}
This material is based upon work supported
by the National Science Foundation under Grant No. CNS-2247754, and by a Meta Security Research Award.

%% % use section* for acknowledgment
%% \ifCLASSOPTIONcompsoc
%%   % The Computer Society usually uses the plural form
%%   \section*{Acknowledgments}
%% \else
%%   % regular IEEE prefers the singular form
%%   \section*{Acknowledgment}
%% \fi

%% The authors would like to thank...



%\bibliographystyle{IEEEtran}
%\bibliography{references}
%% \begin{thebibliography}{1}

%% \bibitem{IEEEhowto:kopka}
%% H.~Kopka and P.~W. Daly, \emph{A Guide to \LaTeX}, 3rd~ed.\hskip 1em plus
%%   0.5em minus 0.4em\relax Harlow, England: Addison-Wesley, 1999.

%% \end{thebibliography}

\section{Appendix}

\subsection{Determining the Stall Bound}
\begin{figure}[h]
  \centering
    \includegraphics[width=\columnwidth,trim = 0 20 0 55,clip]{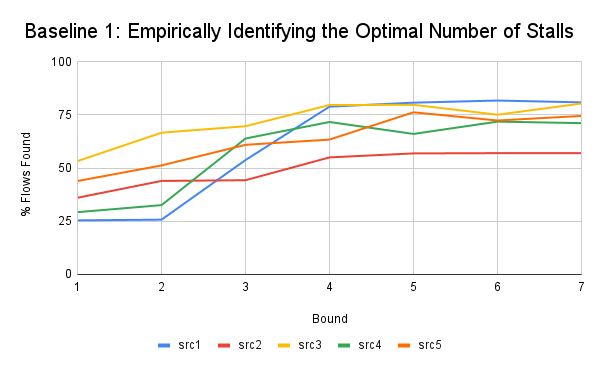}
\caption{Empirically determining the stall
  bound for baseline 1.}
    \label{fig:baseline1-graph-a}
\end{figure}

\begin{figure}[h]
  \centering
    \includegraphics[width=\columnwidth,trim = 0 20 0 65,clip]{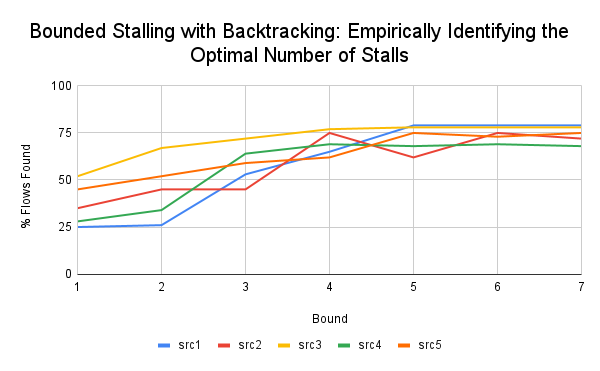}
\caption{Empirically determining the stall
  bound for bounded stalling with backtracking.}
    \label{fig:bd-stalling-graph-a}
\end{figure}

\begin{figure}[h]
  \centering
    \includegraphics[width=\columnwidth,trim = 0 20 0 65,clip]{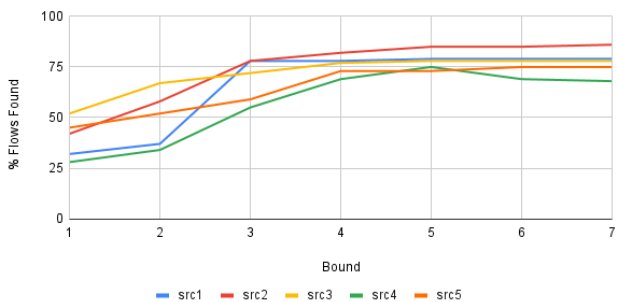}
\caption{Empirically determining the stall
  bound for stalling with the UNSAT core heuristic.}
    \label{fig:stalling-unsat-graph-a}
\end{figure}

We empirically determined the number of clock cycles to stall for each heuristic that involved stalling: baseline 1, 
bounded stalling with backtracking and the UNSAT core heuristic. Once we do not see any significant gains from
stalling for an additional clock cycle, we have found our bound. The results for each of the three heuristics 
are shown in Figures \ref{fig:baseline1-graph-a}, \ref{fig:bd-stalling-graph-a}, and \ref{fig:stalling-unsat-graph-a}.

\end{document}